\documentclass[aps,pra, twocolumn, floatfix]{revtex4-1}

\usepackage{graphicx}
\usepackage{hyperref}
\usepackage{amsmath}
\usepackage{color}
\usepackage{dcolumn}
\usepackage{bm}
\usepackage{float}
\usepackage[normalem]{ulem}
\usepackage{braket}
\usepackage[subrefformat=parens,labelformat=parens,caption=false]{subfig}
\usepackage{bbm}
\usepackage[dvipsnames]{xcolor}
\usepackage{esint}
\usepackage{lineno}
\usepackage{verbatim}
\usepackage{bm}

\usepackage{soul}

\begin{document}
\title{Reply to ``Comment on `Noise, not squeezing, boosts synchronization in the deep quantum regime' "}

\author{W. K. Mok$^{1}$, L.~C.\ Kwek$^{1,2,3,4}$, H.\ Heimonen$^{1}$}
\affiliation{$^{1}$Centre for Quantum Technologies, National University of Singapore, 3 Science Drive 2, Singapore 117543}
\affiliation{$^{2}$Institute of Advanced Studies, Nanyang Technological University, 60 Nanyang View, Singapore 639673}
\affiliation{$^{3}$National Institute of Education, Nanyang Technological University, 1 Nanyang Walk, Singapore 637616}
\affiliation{$^{4}$MajuLab, CNRS-UNS-NUS-NTU International Joint Research Unit, UMI 3654, Singapore}

\begin{abstract}

In this reply we clarify the main points of our manuscript and respond to the critique in the Comment \cite{2002.11514}. In particular, we emphasize that our conclusion ``squeezing loses effectiveness in the deep quantum regime" does not contradict the previous work \cite{PhysRevLett.120.163601} but instead adds to it, and raises fundamental questions on classifying parameter regimes for quantum synchronization. Moreover, we address the concern brought up on the validity of the master equation in the deep quantum regime, and show that our noise-enhanced synchronization differs from previous literature. Through numerical examples, we also demonstrate that the choice of ansatz, while appearing inconsistent, does not lead to erroneous conclusions. Lastly, we expound on the physics of noise-boosted synchronization, and show that it is indeed a genuine feature unique to the deep quantum regime. However, we note that single photon dissipation is a more accurate term, and will move to using that. 

\end{abstract}

\date{\today}
\maketitle

\section{Validity of master equation}

The quantum van der Pol (qvdP) can be described by the master equation (in the weakly nonlinear regime)

	\begin{equation}
\dot{\rho} = -i [H, \rho] + \gamma_1 \mathcal{D}[a^\dag]\rho + \gamma_2 \mathcal{D}[a^2]\rho + \kappa \mathcal{D}[a]\rho
	\label{eq:mastereqn}
	\end{equation}

where $H = \delta a^\dag a + \Omega (a + a^\dag) + \eta (a^2 + a^{\dag 2})$. The standard qvdP does not include the $\kappa \mathcal{D}[a]\rho$ term; here we added it to model the spontaneous decay of the oscillator as a noise term, which is unavoidable in practice.

In the Comment \cite{2002.11514}, the authors question the validity of the quantum van der Pol (qvdP) master equation in the deep quantum limit as we defined in \cite{2002.07488}. The authors claim that the deep quantum limit is in violation of the van Hove limit, which demands $\lambda^2 t \to \text{constant}$ as $t \to \infty$, where $\lambda$ is related to the system-bath coupling strength introduced in the microscopic derivation of the master equation. We present arguments why the equation is valid in the deep quantum regime.

We agree with the requirements of Born-Markov and RWA approximations, as implied by the van Hove limit, and that is why the limit $\gamma_2 / \gamma_1 \to \infty$ is to be taken by $\gamma_1 \to 0$. This allows for making all the approximations appropriately, without going into a strongly coupled regime. One can certainly require both couplings $\gamma_2$ and $\gamma_1$ to be weak with respect to the system frequency $\omega_0$ (baths weakly coupled to the system, in accordance with the van Hove limit) while maintaining a high ratio of $\gamma_2 / \gamma_1$, especially since the hot and cold baths can be regarded as independent baths \cite{1711.07376}. It is thus important to not conflate the high $\gamma_2 / \gamma_1$ ratio with a strong system-bath coupling.

Practically, we note that the limit $\gamma_2 / \gamma_1 \to \infty$ is of course unattainable in any experimental realisation and is only to be regarded as a mathematical limit to simplify the calculations. The concern about the validity of the limit is of academic nature, because the phenomena presented in our manuscript are already evident at $\gamma_2 / \gamma_1 \approx 10 - 100$. We also think that it is inaccurate to claim that ``authors previously stayed away from large values of $\gamma_2 / \gamma_1$", given that several pioneering works in the field use precisely this condition to perturbatively solve the master equation \cite{PhysRevLett.112.094102, doi:10.1002/andp.201400144}. 

Turning the question on its head, suppose that $\gamma_2 / \gamma_1 \to \infty$ results in an invalid master equation. By the same argument, the opposite limit $\gamma_1 / \gamma_2 \to \infty$ (classical limit) would also not be right. Thus, the commenters' seem to imply that the master equation in Eq. (\ref{eq:mastereqn}) does not work in the classical limit. As the master equation is defined as the `quantum van der Pol' oscillator based on this limit, the argument would have wide implications for the whole field. As one can easily check, simulating the master equation for $\gamma_1 \gg \gamma_2$ recovers well-known classical results. We however are now aware that there are many subtle points to consider when discussing any master equation, and it is a worthy subject to study in the future.

\section{Response to Claim I}

In the Comment, the authors remarked that the title of our manuscript is ``scientifically misleading and a serious misrepresentation". In the same paragraph, it was also mentioned that ``the claim that squeezing is ineffective in this regime does not imply that it is ineffective in the original regime $\gamma_2 / \gamma_1 \approx 3$ considered \cite{2002.11514}." Unfortunately, this appears to be a straw man argument, for it was never claimed in our manuscript that the results of \cite{PhysRevLett.120.163601} were wrong. In fact, as has been shown prominently in Figs. 4(d) and 4(e), and further emphasized in the Conclusion of \cite{2002.07488}, that squeezing indeed results in strong frequency entrainment at $\gamma_2 / \gamma_1 = 3$ and outperforms harmonic driving. However, as we found out, the entrainment effect of squeezing weakens drastically with increasing $\gamma_2 / \gamma_1$, while the entrainment due to harmonic driving remains robust. This is why the title of the paper emphasizes the deep quantum regime. The crossover point where harmonic driving begins to outperform squeezing was found to be at approximately $\gamma_2 / \gamma_1 \approx 13$. A literature search \cite{lee2013quantum,PhysRevLett.112.094102,PhysRevLett.120.163601,PhysRevLett.121.053601,doi:10.1002/andp.201400144, ameri2015mutual,morgan2015oscillation,ishibashi2017oscillation,koppenhofer2019optimal,
lorch2016genuine,lee2014entanglement,weiss2017quantum,dutta2019critical,es2019synchronization}  shows that the phrases "deep quantum regime", "deeply quantum regime", and "quantum regime" have been used interchangeably to refer to various values of $\gamma_2 / \gamma_1 $ from 1 to 1000 and $\infty$. There is no clear consensus as to what values the phrases refer to. It is precisely for this reason that the need to classify the parameter regimes (in Tab. I of \cite{2002.07488}) into ``quantum regime" and``deep quantum regime" arises. In our definition of the deep quantum regime (as rightly pointed out in the Comment) the population of the state $\ket{2}$ becomes so negligible such that squeezing does not produce a strong observable synchronization effect. It is also in this regime where the effect of our single photon noise on the vdP in fact improves its synchronization to the drive, instead of destroying the synchronization due to decoherence as one might expect. We hope that this clarifies the title of our manuscript. However, for the next revision of the paper we will move to the title `Single photon dissipation boosts synchronisation in the deep quantum regime'. 

The authors of the comment also say that it is clear from the form of the steady state of Eq. \ref{eq:mastereqn} ($\rho_{ss} = \frac{2}{3}\ket{0}\bra{0}+\frac{1}{3}\ket{1}\bra{1}$) that squeezing does not work in the deep quantum limit and that is the very reason in the paper \cite{PhysRevLett.120.163601} they study squeezing at $\gamma_2 / \gamma_1 = 3$. This point is also not made in \citep{PhysRevLett.120.163601} and it leads the reader to believe that squeezing always enhances quantum synchronization. This is the reason for mentioning it in our title. We would like to highlight that the steady state is derived in exactly the same limit ($\gamma_2 / \gamma_1 \to \infty$) in which the authors claimed the master equation is invalid. Our results cannot be both trivially true and false at the same time.

\section{Response to Claim II}

The second claim of the comment is that our findings regarding noise boosting synchronization in the deep quantum regime are not novel. They cite two works \citep{PhysRevE.101.020201, PhysRevLett.121.053601} as examples where they claim a similar noise boost has been seen. While we agree with the commenters' that we consider a specific type of noise, single photon dissipation, we beg to differ that similar effects have studied in the reference. Neither system's dissipators can be mapped onto the qvdP oscillator to the best of our knowledge, separating our work from them. We now discuss the two papers individually and highlight why they do not show analogous physics either. However, in hindsight we agree that the term noise is too general to refer to this specific process, and we will amend the manuscript and its title regarding this.

\subsection{The Qutrit} 
In the manuscript \cite{PhysRevLett.121.053601} a limit cycle for a qutrit is derived and studied. The authors of the comment \citep{2002.11514} state that because the system cannot be synchronized for equal dissipation rates $\gamma_d$ and $\gamma_g$, this is a discovery of a noise boost of synchronization. The authors of the original study do not call changing the dissipation rates noise, and it is not accurate to do so either. The two dissipators with the rates $\gamma_d$ and $\gamma_g$ are responsible for creating the limit cycle, and the rates represent system-bath coupling strengths. Varying the coupling strengths does not have a clear meaning as introducing noise to the system. We emphasize that the noise-enhanced synchronization in our manuscript does not come from manipulating the rates which are responsible for constructing the original limit cycle, but rather from external noise on an existing limit cycle which can be thought of as an imperfection (possibly due to experimental limitations). In analogy, let us consider the qvdP limit cycle which is created from the balance between the single-photon pumping and two-photon loss. The amplitude $N_0$ is purely governed by $\gamma_1$ and $\gamma_2$. Now, by increasing $\gamma_1$ with respect to $\gamma_2$, it should not be surprising that one can observe a `synchronization boost', because adding more pumping pushes the oscillator into higher energy states, which becomes less affected by quantum fluctuations, well-known to inhibit synchronization \cite{PhysRevLett.112.094102}.

\subsection{The Heat Engine} 
The nanoscale heat engine in \cite{PhysRevE.101.020201} has been shown to exhibit different amounts of synchronization for varying ratios of bath temperatures $T_c/T_h$ in Fig. 2(c) of the paper. In the paper a qutrit is coupled to two heat baths, so the concept of thermal noise is clear. However, what Fig. 2(c) shows is not the effect of this noise. While we can associate thermal noise to a non-zero absolute temperature, the plots show the effect of varying relative temperatures. The same ratios (and therefore amount of synchronization) can be achieved by vastly different absolute temperatures in each bath (and therefore with very different amounts of thermal noise present). Therefore the results in our manuscript, where we add single photon losses, have nothing to do with the results found for the heat engine.

Thus, we stress that our work is very different from both of the examples above and represents a truly novel phenomenon. Our noise-enhanced synchronization is a genuine quantum effect which appears only in the regime of large $\gamma_2 / \gamma_1$, and is not possible to be observed in either the semiclassical regime or the classical vdP (as shown in the Appendix of \cite{2002.07488}). 

\section{Response to ``Ansatz inconsistency"}

It was highlighted in the Comment that the coherence $|\rho_{02}|$ (we believe the authors' meant $|\rho_{12}|$ instead) is almost twice the value of the population of $\ket{2}$, $|\rho_{22}|$, in the deep quantum regime. This is a problem because it is inconsistent to neglect $\rho_{02}$ and $\rho_{12}$ while keeping $\rho_{22}$ in the density matrix ansatz. The element $\rho_{22}$ is required due to the $\mathcal{D}[a^2]\rho$ dissipator in the master equation. 

We appreciate that the commenters have carefully read our manuscript and pointed this out. While we agree that strictly speaking this treatment is inconsistent, there are several reasons why it does not diminish our analysis. The dropped terms are all coherences, which will mainly affect the calculation of the synchronization measure. Our synchronization measure is given by the MRL $| \exp(i \hat{\phi})| = | \rho_{01} + \rho_{12} + \rho_{23} + ... |$, which does not include $\rho_{02}$. Dropping $\rho_{12}$ simplifies the calculations considerably, while not introducing too much error as shown in the comment Fig. 1(a) (and $|\rho_{12}| \ll |\rho_{01}|$). Moreover, as evident in Fig. 2(b) of our manuscript \cite{2002.07488}, the analytical approximation obtained from the ansatz agrees very well with numerical results. The appendix of the manuscript provides a comparison between the analytics and numerics. The authors in \cite{2002.11514} correctly mentioned that the additional coherences will ``increase any reasonable measure of synchronization". All the data in the manuscript are obtained from numerical simulations (unless explicitly stated otherwise), with a much higher Hilbert space truncation than the minimum given in the ansatz. This includes the two main results regarding squeezing and the effects of noise. No redoing of the analysis is thus necessary.
   
\section{Response to ``Physics behind noise-boosted synchronization"}

The comment highlights that we ``fail to discuss the fact that noise boosts synchronization only for a very limited range of $\kappa$". They offer the following explanation to the decrease in the boost:``while bringing down population from the first excited level to the ground level works for a little bit, the deleterious effect of a linear damp on the populations overwhelms any advantage provided to the coherence". Our understanding of the weakening of the noise boost is the same as that of the commenters'. We thank the commenters for pointing out that we have mistakenly taken the explanation out of the manuscript, and will update this into the next version. 

To further comment on the phenomenon, we would like to highlight that in the limit of $\kappa \gg \Omega$, the timescale in which the the damping $\kappa^{-1}$ occurs, is much faster than that of the driving (with timescale $\Omega^{-1}$) such that it becomes much harder for coherences to build up. In that case, the noise term indeed becomes detrimental to synchronization. A similar plot to Fig. 1(b) of the Comment was also plotted in Fig. 4(b) of \cite{2002.07488}, showing exactly the trade off between noise-enhanced synchronization and excessive spontaneous decay. We note that a similar matching of time-scales is also required for other noise-assisted processes such as stochastic resonance \cite{RevModPhys.70.223}.

\section{Response to ``Limit cycle size"}

In the comment the authors' disagree with studying the oscillator along threshold driving conditions and claim that the increasing driving could be behind the noise boost of synchronization.

In our appendix in \cite{2002.07488} we present a mathematical argument for the condition for the noise boost:
	\begin{equation}
\frac{M^\prime}{M} > \frac{D^\prime}{D}
	\label{eq:condition}
	\end{equation}
where
	\begin{equation}
M = 2 \Omega [\gamma_1 (\gamma_2 - \kappa) + \kappa (\gamma_2 + \kappa)] \sqrt{4\delta^2 + (\kappa + 3\gamma_1)^2}
	\end{equation}
and 
	\begin{equation}
	\begin{split}
D &= \gamma _1 [4 \gamma _1 \left(\delta ^2+4 \kappa ^2+3 \Omega ^2\right)+15 \gamma _1^2 \kappa +9 \gamma _1^3+4 \delta ^2 \kappa \\
&+7 \kappa  \left(\kappa ^2+4 \Omega ^2\right)] +\kappa ^2 \left(4 \delta ^2+\kappa ^2+8 \Omega ^2\right) \\
&+\gamma _2 \left(3 \gamma _1+\kappa \right) \left(6 \gamma _1 \kappa +9 \gamma _1^2+4 \delta ^2+\kappa ^2+8 \Omega ^2\right)
	\end{split}
	\end{equation}
(with $M^\prime \equiv \partial_\kappa M (\kappa = 0)$ and similar definition for $D^\prime$) without assuming the deep quantum limit. When this condition is met, it is equivalent to $\partial_\kappa S (\kappa = 0) > 0$ which implies that noise improves synchronization, ceteris paribus. It is straightforward to verify that the condition given in Fig. 4(b) of our manuscript satisfies Eq. (\ref{eq:condition}) which is consistent with the notion that noise boosts synchronization. If we use the simplified result in the deep quantum limit, then it is also shown in the appendix that $\partial_\kappa S (\kappa = 0) > 0$ is always true. This conclusion holds for all $\Omega > 0$, and provides mathematical justification for our claim that noise enhances synchronization in the deep quantum regime.

Thanks to the comment, we noticed that it was incorrectly mentioned in our manuscript that the qvdP shown in Figs. 4(a) and 4(b) are driven at threshold. Instead the driving in the figures is fixed at $\Omega/\gamma_1 = 1$. This error will be rectified post-haste, and it is unfortunate to have resulted in the misinterpretation in the Comment. 

Since the figures are in fact showing the synchronization as a function of noise only (constant driving), the issue mentioned in the comment should be cleared. Classical intuition however suggests that threshold driving would be the correct way to plot the curve. Both results can be seen in Fig 3, where following the white line is along threshold driving and any horizontal cross-section corresponds to pure addition of noise. The noise boost can be seen along both trajectories.

\section{Conclusion}

We thank the commenters' on their careful reading of our manuscript and for raising their worries regarding it. We reject the major criticisms in the comment and hope that the above arguments clarify our stance. This reply explained the need for classifying regimes of the qvdP and why our phenomena are new and interesting for the field. We will move to fix the minor mistakes and omissions pointed out by the commenters in an update to the article. We will also move to using the more accurate term `single-photon dissipation', instead of the general `noise', and to the title `Single photon dissipation boosts synchronisation in the deep quantum regime'. 

\bibliographystyle{apsrev4-1}
\bibliography{qsync_reply}

\end{document}